\documentclass[twocolumn,aps,pra,superscriptaddress]{revtex4-1}

\usepackage{amsmath}
\usepackage{amssymb}
\usepackage{amsfonts}
\usepackage{array}
\usepackage[english]{babel}
\usepackage{bm}
\usepackage{color}
\usepackage{floatrow}
\usepackage{graphicx}
\usepackage{hyperref}
\hypersetup{colorlinks=true, urlcolor=blue, citecolor=blue, linkcolor=red}
\usepackage[label font={bf}]{subfig}

\begin{document}

\title{Combined effect of mutually frequency-detuned strong and weak drives on a two-level system: Envelope of the Rabi oscillations}

 \author{M. E. Raikh}
\affiliation{Department of Physics and
Astronomy, University of Utah, Salt Lake City, UT 84112}

\begin{abstract} 
 Near-resonant ac-drive acting on a two-level system induces the Rabi oscillations of the level occupations. It is shown that additional weak drive properly frequency-detuned from the primary drive causes a resonant response. This response manifests itself in the emergence of the envelope of the oscillations. At resonance, the inverse period of the envelope is proportional 
 to the {\em amplitude} of the
 weak drive. The resonant condition reads: difference of frequencies between the two drives is equal to the ac-splitting of quasilevels 
 in the field of the strong drive. 
  Technically, the resonance can be inferred from the analogy 
  between the equations for the time-evolution of the
 spin amplitude and the 
 Mathieu equation, which describes e.g. the parametric resonance.

\end{abstract}

\maketitle

\section{Introduction}

Dynamics of a spin placed in  a magnetic field, ${\bf B}={\bf z_0}B$, and subjected  to a sinusoidal ac drive, ${\bf b}
=2b{\bf x}_0\cos\omega t $, is known in tiniest details since the celebrated papers [\onlinecite{Rabi,Bloch,Autler-Townes, Shirley,Mollow}]. Most prominent is the situation when $\omega$ is close to the
Zeeman splitting, $\Delta_Z = \gamma B$, of the spin levels. Here $\gamma$ is the gyromagnetic ratio. Then the probabilities of $\uparrow$ and $\downarrow$ spin projections 
 oscillate with the generalized Rabi frequency
\begin{equation}
\label{Rabi}    
\Theta_R=\Bigl[\left(\Delta_Z-\omega\right)^2 
+\Omega_R^2\Bigr]^{1/2},
\end{equation}
where $\Omega_R=\gamma b$. Concerning the $x$ and 
$y$ spin projections, their dynamics comprises three
frequencies:\cite{Mollow} $\omega$, and $\omega\pm \Theta_R$ (Mollow triplet). 

Four underlying frequencies of the spin dynamics suggest the way in which the driven spin can be manipulated by the {\em secondary} ac field. Namely,
the frequency of the secondary field, $\omega_m$, can either be low, of the order of 
$\Omega_R \ll \omega$ (see e.g. 
Refs. [\onlinecite{Fedaruk1,Fedaruk2,Fedaruk3,Fedaruk4,Glenn}]), or it can be high\cite{THEORY,WeakAndStrong,Experiment1,Experiment2,Experiment3,EXPERIMENTcold,TwoDrives,numeric,numeric1} 
and detuned from $\omega$  approximately by $\Omega_R$. Within the
first technique,  the modification of the primary Rabi oscillations by the
secondary drive depends strongly on the relation between the magnitude and
frequency of the secondary drive. In particular, strong secondary drive slows down the Rabi oscillations\cite{Glenn}. When the secondary drive is weak, it can affect the Rabi oscillations only under the resonance condition $\omega_m\approx \Omega_R$. Under this condition, primary oscillations develop an 
envelope.\cite{Fedaruk3,Glenn} In the language of quasienergies this effect can be
interpreted as follows. While the primary drive causes the ac-splitting of the bare levels into the quasilevels, under the resonant secondary drive each of these quasilevels gets split additionally.   

Within the second technique, the effect of 
two near-resonant drives on the two-level system is much more dramatic.
Theoretical studies\cite{THEORY,WeakAndStrong,TwoDrives,numeric,numeric1}  
indicate that adding a weak second (probe) drive can covert the regime of the Rabi oscillations induced by the primary drive (pump) into a chaotic 
behavior of the observables. The underlying reason for chaos is that, 
in the presence of two drives, the Floquet theorem is lifted (if the pump
and probe frequencies are incommensurate).
Meanwhile, the behavior of fluorescence observed in experiments on bichromatically driven two-level systems\cite{Experiment1,Experiment2,Experiment3,EXPERIMENTcold} 
indicate that the effect of two drives is much less dramatic than in theory.
In the time domain, the probe drive leads to the additional modulation of the Rabi oscillations. In the frequency domain, the pump leads to the splitting
of each individual peak of the Mollow triplet into the daughter triplets. 

In the theoretical papers on bichromatic drive  
the results 
are presented in the form of numerical curves calculated for certain sets of parameters. This is certainly justified for revealing of  chaos. But in the regimes when the effect of secondary drive is modest, there are no analytical formulas predicting the modified behavior of the
Rabi oscillations for given amplitudes and frequencies of two drives. 
Such formulas are derived in the present paper. The reason why these formulas
can be derived is that, within the rotating wave approximation (RWA), the Floquet theorem
applies not for the primary drive frequency but rather for the {\em difference}  between the frequencies of two drives. 
Our analytical treatment yields the resonance condition for which the  effect of a weak  secondary drive on the Rabi oscillations is most
pronounced.

\begin{figure}
\label{F1}
\includegraphics[scale=0.3]{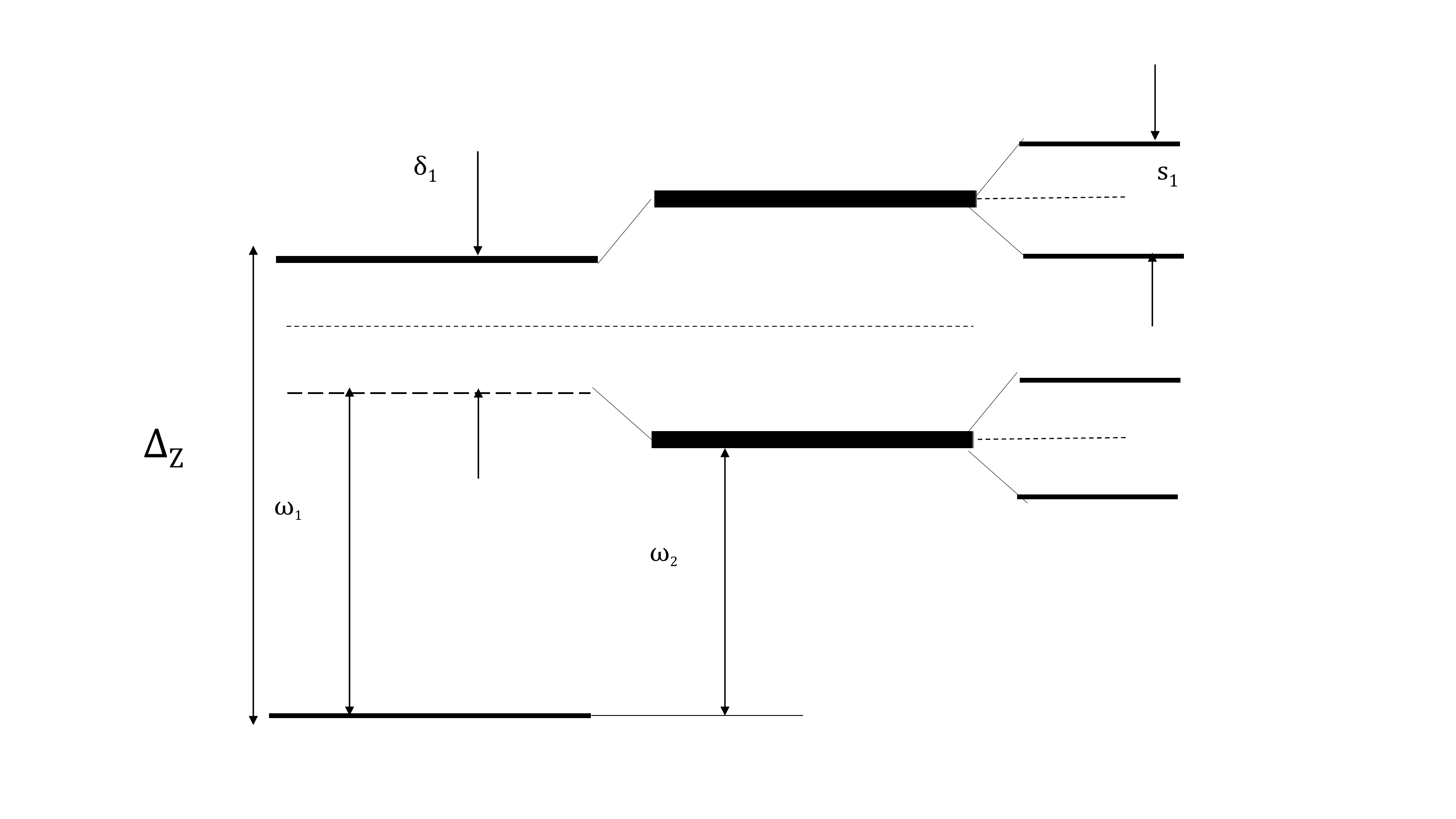}
\caption{(Color online) Schematic illustration of a
two-level system in a magnetic field driven by the
ac fields with frequencies $\omega_1$ and $\omega_2$
close to the level separation, $\Delta_Z$.
Strong drive with magnitude, $\lambda_1$, and 
ac-splitting of the Zeeman levels, which manifests
itself in the Rabi oscillations of the level 
$\lambda_2\ll \lambda_1$ properly detuned from the
primary drive, see Eq. \ref{Bragg}, leads to an
additional splitting, $s_1$, of the ac-split levels,
which, in turn, results in the envelope of the Rabi
oscillations with a period $\frac{2\pi}{s_1}$.}
\end{figure}

\section{Evolution of the amplitudes of the spin components}

Denote with $a_1$ and $a_2$ the amplitudes of $\uparrow$ and $\downarrow$ spin 
components. In the presence of two drives these amplitudes evolve with time as
\begin{align}
&i\frac{da_1}{dt}=\frac{\Delta_Z}{2}a_1+\left[\lambda_1e^{-i\omega_1t}+\lambda_2
e^{-i\omega_2t}\right]a_2, \label{RWA1}\\
&i\frac{da_2}{dt}=-\frac{\Delta_Z}{2}a_2+\left[\lambda_1e^{i\omega_1t}+\lambda_2
e^{i\omega_2t}\right]a_1, \label{RWA2}
\end{align}
where $\omega_1$ and $\omega_2$ are  the frequencies of the primary and secondary drives, respectively, while $\lambda_1$ and $\lambda_2$ are the amplitudes of these drives in the frequency units. For simplicity we neglect the phase  shift between the two drives, and, thus, choose $\lambda_1$, $\lambda_2$ to be real. We also assume that both drives are weak, namely,
$\lambda_1, \lambda_2 \ll~\Delta_Z$. This justifies the replacement of $\cos \omega_1t$ and $\cos \omega_2t$ in Eqs. \ref{RWA1} and \ref{RWA2} by the exponents, which is the essence of RWA. Incorporating of counter-rotating exponents
amounts to renormalization of $\Delta_Z$ by the Bloch-Siegert shift\cite{Bloch} of
the order of 
$\frac{\lambda_1^2}{\Delta_Z}, \frac{\lambda_2^2}{\Delta_Z} \ll \Delta_Z$.
Following the standard procedure for a single-frequency drive, we introduce new variables

\begin{equation}
\label{variables}
A_1=a_1\exp\left(\frac{i\Delta_Z}{2}t   \right),~~A_2=a_2\exp\left(-\frac{i\Delta_Z}{2}t\right),
\end{equation}
which allows to exclude high frequency, $\Delta_Z$, from the system Eqs. \ref{RWA1} and \ref{RWA2}, which takes the form
\begin{align}
& i\frac{dA_1}{dt}=\left[\lambda_1e^{-i\delta_1t}+\lambda_2
e^{-i\delta_2t}\right]A_2, \label{A1}\\
& i\frac{dA_2}{dt}=\left[\lambda_1e^{i\delta_1t}+\lambda_2
e^{i\delta_2t}\right]A_1, \label{A2}
\end{align}
where the detunings $\delta_1$ and $\delta_2$ are defined as
\begin{equation}
\label{detunings}
\delta_1=\Delta_Z-\omega_1,~~\delta_2=\Delta_Z-\omega_2.
\end{equation} 
Next we reduce the system Eqs. \ref{A1}, \ref{A2} to a single
second-order differential equation. Expressing $A_2$ from Eq. \ref{A1}
and substituting into Eq. \ref{A2}, we get
\begin{align}
\label{second}
&-\frac{d^2A_1}{dt^2}-if(t)\frac{dA_1}{dt}\nonumber\\
& =\left[\lambda_1^2 +\lambda_2^2   +\lambda_1\lambda_2e^{i(\delta_1-\delta_2)t}    +\lambda_2\lambda_1
e^{-i(\delta_1-\delta_2)t}\right]A_1.
\end{align}
where the function $f(t)$ is given by
\begin{align}
\label{f}
&f(t)=\frac{\lambda_1\delta_1e^{i\delta_1t}
+\lambda_2\delta_2e^{i\delta_2t}}{\lambda_1e^{i\delta_1t}+\lambda_2e^{i\delta_2t} }\nonumber\\
&=\frac{\delta_1 +\delta_2}{2}+\frac{\delta_1-\delta_2}{2}
\Biggl(\frac{1-\frac{\lambda_2}{\lambda_1}e^{i(\delta_2-\delta_1)t}}{1+\frac{\lambda_2}{\lambda_1}e^{i(\delta_2-\delta_1)t}}      \Biggr).
\end{align}
It is seen from the expression Eq. \ref{f} that the function $f(t)$
is periodic with a period $2\pi/(\delta_2-\delta_1)$. Equally, the right-hand side of Eq. \ref{second} is periodic with the same period.
 We thus conclude that the Floquet theorem applies to Eq.~ (\ref{second}). To boost the similarity to the Mathieu equation, we exclude the first derivative from Eq. (\ref{second}) by introducing the new variable 
 
\begin{align}
\label{tilde}
& {\tilde A}_1(t) = A_1(t)\exp\Biggl[\frac{i}{2}\int_{0}^t
 dt'f(t')\Biggr]\nonumber\\
 &=A_1(t)\Biggl[1+\frac{\lambda_2}{\lambda_1}e^{i(\delta_2-\delta_1)t }    \Biggr]^{1/2}\exp\Biggl({\frac{i\delta_1 t}{2}}\Biggr).
\end{align}
 Substituting Eq. \ref{tilde} into Eq. \ref{second}
 we arrive to the following equation for ${\tilde A}_1$
\begin{align}
\label{TILDE}
&-\frac{d^2{{\tilde A}_1}}{dt^2}
-\frac{{\tilde A}_1}{4}
\Biggl\{\delta_1^2+\delta_2^2           -\Biggl( \frac{\lambda_2\delta_1e^{i\delta_2t}
+\lambda_1\delta_2e^{i\delta_1t}}
{\lambda_2e^{i\delta_2t}
+\lambda_1e^{i\delta_1t}}
\Biggr)^2\Biggr\}\nonumber\\
&=\left[\lambda_1^2 +\lambda_2^2   +\lambda_1\lambda_2e^{i(\delta_1-\delta_2)t}    +\lambda_2\lambda_1
e^{-i(\delta_1-\delta_2)t}\right]{\tilde A}_1.
\end{align}
Note that in the limit when the secondary drive is absent, $\lambda_2=0$,
Eq.~ \ref{TILDE} reproduces the Rabi oscillations in the field of the primary drive, namely, the solutions of Eq.~ \ref{TILDE} have the form 
${\tilde A}_1\propto \exp\left[\pm i\left( \lambda_1^2+\frac{\delta_1^2}{4}       \right)^{1/2}t\right].$

\vspace{3mm}

\section{Weak secondary drive}

Within the adopted RWA approximation Eq. 
\ref{TILDE} applies for arbitrary relation between two drives $\lambda_1$ and $\lambda_2$
and arbitrary relation between the detunings
$\delta_1$ and $\delta_2$ as long as they are much smaller than $\omega$. Next we take the limit of a weak secondary drive  
$\lambda_2 \ll \lambda_1$. In the lowest order in the secondary drive we can neglect the
term $\lambda_2^2$ in the right-hand side. Concerning the
left-hand side, we expand the
ratio in the brackets as follows

\begin{align}
\label{expand}
& \frac{\lambda_1\delta_2e^{i\delta_1 t} +    \lambda_2\delta_1e^{i\delta_2 t}}
{\lambda_1e^{i\delta_1 t} +\lambda_2e^{i\delta_2 t} }\nonumber\\
& \approx\delta_2\Biggl[1+\frac{\lambda_2}{\lambda_1}\Bigl(\frac{\delta_1}{\delta_2} -1  \Bigr)e^{i(\delta_2-\delta_1) t}      \Biggr].
\end{align}
With the above simplifications Eq. \ref{TILDE}
takes the form 
\begin{align}
\label{simple}
&-\frac{d^2{{\tilde A}_1}}{dt^2} - 
\Biggl[\lambda_1^2+\frac{\delta_1^2}{4}      +\frac{\lambda_2}{2\lambda_1}\delta_2(\delta_1-
\delta_2)e^{-i(\delta_1-\delta_2)t}\Biggr]{\tilde A}_1
\nonumber\\
&=\left[\lambda_1\lambda_2e^{i(\delta_1-\delta_2)t}    +\lambda_2\lambda_1
e^{-i(\delta_1-\delta_2)t}\right]{\tilde A}_1.
\end{align}
There are two terms 
proportional to $\lambda_2$ in
Eq. \ref{simple}. Assuming that $\delta_1$ and $\delta_2$ are of the same order, the term in the left-hand side can be estimated as
$\lambda_2\delta_1^2/\lambda_1$, while the term in the right-hand side
is of the order of $\lambda_2\lambda_1$. If $\lambda_1$  is much greater than $\delta_1$, which corresponds to the limit of developed Rabi oscillations, the  term in the right-hand side dominates. Note now that,
if one neglects the term in the left-hand side, Eq. \ref{simple} reduces
to the classical Mathieu equation, which describes e.g.  the electron motion in a weak one-dimensional potential. The time in Eq. \ref{simple}  plays the role of coordinate. The sum $\lambda_1^2+\frac{\delta_1^2}{4}$
plays the role of energy. Despite the similarity, there is a dramatic  difference
between Eq. \ref{simple} and the Mathieu equation, since the 
Mathieu equation describes  the "forbidden gaps" or the domains 
of instabilities, while Eq.~\ref{simple} does not. Still, as we will see below, the position of resonance captured by Eq. \ref{simple} can be found from the same "Bragg condition" as the
position of the center of the forbidden gap, namely

\begin{equation}
\label{Bragg}
 2\Biggl( \lambda_1^2+\frac{\delta_1^2}{4}  \Biggr)^{1/2}=|\delta_1-\delta_2|.
\end{equation}

To trace how the resonance Eq. \ref{Bragg} emerges, we search for
the solution of Eq. \ref{simple} in the Floquet form
\begin{equation}
 \label{form}  
 {\tilde A}_1(t)=f\exp{\{ist\}} +g\exp{\Big\{i(s+\delta_1-\delta_2)t\Big\}}.
\end{equation}

Substituting this form into Eq. \ref{simple} and equating the
coefficients in front of the two exponents, we arrive to  the system
\begin{align}
\label{AA}
&\left(\lambda_1^2 +\frac{\delta_1^2}{4}-s^2\right)f=
-\lambda_1\lambda_2\Bigg[\frac{\delta_2(\delta_1-\delta_2)}
{2\lambda_1^2}+1\Bigg]g, \nonumber\\  
&\left(\lambda_1^2 +\frac{\delta_1^2}{4}-(s+\delta_1-\delta_2)^2\right)g=-\lambda_1\lambda_2f. 
\end{align}
The brackets in the left-hand sides  of the two equations coincide
under the condition: $s+\delta_1-\delta_2=-s$. Thus we set
\begin{equation}
\label{set}
s=\frac{\delta_2-\delta_1}{2} +s_1,   
\end{equation}
and assume that $s_1\ll |\delta_2-\delta_1|$. Upon neglecting the
$s_1^2$-term, the system Eq. \ref{AA} takes the form
\begin{align}
\label{AAA}
 &\Bigg[\lambda_1^2 -\frac{\delta_2^2-2\delta_2\delta_1}{4}-s_1(\delta_2-\delta_1)\Bigg]f=\nonumber\\
 &
-\lambda_1\lambda_2\Bigg[\frac{\delta_2(\delta_1-\delta_2)}
{2\lambda_1^2}+1\Bigg]g, \nonumber\\  
& \Bigg[\lambda_1^2 -\frac{\delta_2^2-2\delta_2\delta_1}{4}+s_1(\delta_2-\delta_1)\Bigg]                                        g= \nonumber\\
&-\lambda_1\lambda_2f.    
\end{align}
Finally, upon multiplying the two last equations, we obtain the
expression for the Floquet exponent
\begin{align}
\label{s1}
 &s_1^2\left(\delta_2-\delta_1\right)^2= \nonumber\\
 &\left[\lambda_1^2-\frac{\delta_2^2}{4}+\frac{\delta_1\delta_2}{2} \right]^2-\lambda_2^2\left[\frac{\delta_1\delta_2-\delta_2^2}{2}
 +\lambda_1^2\right].
\end{align}
Note now that the first bracket on the right-hand side of Eq. 
\ref{s1} turns to zero when Eq. \ref{Bragg} is satisfied, i.e. at resonance. It is seen from Eq. \ref{s1} that the resonance manifests
itself via the anomalous sensitivity of the Floquet exponent to the magnitude, $\lambda_2$,  of the secondary drive. It is also 
instructive to rewrite  Eq. \ref{s1} in a different form by using the resonant condition Eq. \ref{Bragg}. Substituting $\lambda_1^2=\frac{1}{4}\left(\delta_2^2-2\delta_1\delta_2\right)$ into Eq. \ref{s1}
we get

\begin{equation}
\label{Floquet}
s_1=\pm \frac{\lambda_2\delta_2}{2(\delta_2-\delta_1)}=\pm \frac{\lambda_2\delta_2}{4\left(\lambda_1^2+\frac{\delta_2^2}{4}    \right)^{1/2}  }.
\end{equation}
In the second identity we have again used the resonant 
condition Eq. \ref{Bragg}.

Yet another way to cast  Eq. \ref{s1} is to express $\delta_2$ from
the resonant condition. This yields
\begin{equation}
\label{Floquet1} 
s_1=\pm \frac{\lambda_2}{4\Theta_R}\left(\Delta_Z-\omega_1 \pm 2\Theta_R   \right).
\end{equation}
In the last identity we have returned to the original notations, generalized Rabi frequency and the detuning of the primary drive
(see Eq. \ref{Rabi}). Since we assumed that $s_1$ is much smaller than
the frequency spacing between the two drives, the criterion of applicability of Eq. \ref{Floquet1} is $\lambda_2 \ll |\delta_2-\delta_1|$.

 Eq. \ref{Floquet1} is our main result. It suggests that, upon tuning the frequency of the secondary drive to the resonance  
Eq.~\ref{Bragg}, the Floquet exponent, describing the drive-induced modification of the Rabi oscillations,  increases {\em linearly} with
the amplitude of the secondary drive.

\vspace{3mm}

\section{Discussion}
({\em i}). Upon increasing of $\lambda_2$  
the spectrum of the secondary Rabi oscillations becomes progressively richer 
due to emergence of the higher-order Bragg resonances for which the detunings satisfy the condition
\begin{equation}
\label{BraggN}
 2\Biggl( \lambda_1^2+\frac{\delta_1^2}{4}  \Biggr)^{1/2}=N|\delta_1-\delta_2|
\end{equation}
with higher $N$. To trace the emergence 
of these resonances, we expand the
ratio Eq. \ref{expand} to the second order
in the ratio $\frac{\lambda_2}{\lambda_1}$
\begin{align}
 \label{second}
& \Biggl( \frac{\lambda_2\delta_1e^{i\delta_2t}
+\lambda_1\delta_2e^{i\delta_1t}}
{\lambda_2e^{i\delta_2t}
+\lambda_1e^{i\delta_1t}}
\Biggr)^2
\approx \delta_2^2\Biggl[1+2\frac{\lambda_2}{\lambda_1}
\Biggl(\frac{\delta_1}{\delta_2} -1   \Biggr)e^{i(\delta_2-\delta_1)t   }\nonumber\\
&+ \Biggl(\frac{\lambda_2}{\lambda_1}\Biggr)^2
\Biggl(\frac{\delta_1}{\delta_2} -1   \Biggr)
\Biggl(\frac{\delta_1}{\delta_2} -3  \Biggr)e^{2i(\delta_2-\delta_1)t   }
\Biggr].
\end{align}
We see that the new term in the expansion
gives rise to the second harmonics in the "effective potential" which causes the resonance with $N=2$. 

({\em ii}). Floquet exponent translates into an envelope. 
The situation addressed in the present paper is standard and 
was previously addressed in the literature, see e.g.
Refs.~\onlinecite{THEORY,WeakAndStrong,Experiment1,Experiment2,Experiment3,EXPERIMENTcold,TwoDrives,numeric,numeric1}. We considered a 
two-level system driven by two ac fields with strongly different 
amplitudes. A new finding reported in the present paper is the "Bragg resonance" Eq.~ \ref{Bragg} which takes place when the difference of frequencies of the drives is equal to the ac-splitting of  quasienergies
in the field of the strong drive. Interplay of the two drives can be interpreted as
a modulation of the amplitude of a primary drive with a frequency $\omega_1-\omega_2=\delta_2-\delta_1$, which couples the ac-split levels. The result of this 
coupling is the secondary splitting of quasienergies.   Most importantly, at resonance, the magnitude 
of this secondary splitting is {\em linear} in the amplitude of the weak drive.
Emergence of the Bragg resonance
is illustrated schematically in the figure.
Physically, the Bragg resonance manifests itself in the modulation of the primary Rabi oscillations with a frequency, $s_1$, proportional to the amplitude, $\lambda_2$, of the weak
drive, see  Eq. \ref{Floquet1}.     It should be emphasized that the {\em depth} of modulation is {\em independent} of $\lambda_2$. Indeed, this depth is determined
by the interference of the amplitudes $f$
and $g$ in Eq. \ref{form}. At resonance, the
ratio of these amplitudes can be expressed from the system Eq. \ref{AAA} as follows
\begin{equation}
\label{RATIO} 
\frac{f}{g}=-\frac{s_1(\delta_2-\delta_1)}
{\lambda_1\lambda_2}.
\end{equation}
Substituting the Floquet exponent from 
Eq.~\ref{Floquet} we find

\begin{equation}
\label{RATIO1} 
\frac{f}{g}=\pm \frac{\delta_2}{2\lambda_1},
\end{equation}
i.e. the amplitude $\lambda_2$ drops out. In other words, no matter how weak is the secondary drive, at resonance, it lead s to the modulation of the primary Rabi oscillations with a depth $\sim 1$. If the
detuning, $\delta_2$, of the secondary drive
does not satisfy the Bragg condition,
Eq. \ref{Bragg} by a small quantity, 
$\Delta\delta_2$, the modulation of the
Rabi oscillations gradually vanishes. As can be seen
from the system Eq. \ref{AAA}, the Floquet
exponent changes with $\Delta\delta_2$ as
\begin{equation}
\label{Delta}
s_1(\Delta\delta_2)=\Bigl[s_1(0)^2+
\frac{(\Delta\delta_2)^2}{4}\Bigr]^{1/2},
\end{equation}
i.e. similarly to the primary Rabi oscillations. This suggests that the dependence of the Floquet exponent on
the magnitude to the secondary drive gradually transforms from linear to quadratic.

Note that the resonance Eq. \ref{Bragg}
was not uncovered in the previous theoretical studies. The reason
is that these studies attempted to incorporate finite lifetimes
of the levels from the very beginning. Then, the $2\times 2$ system
Eqs. \ref{RWA1}, \ref{RWA2} transforms into the $4\times 4$ system 
for the elements of the density matrix, which necessarily complicates the analytical treatment.

({\em iii}). Clearly, the solution of Eq. \ref{TILDE} exhibits anomalous behavior in a particular case when the amplitudes of
two drives are equal: $\lambda_1=\lambda_2=\lambda$, while their
frequencies are detuned {\em symmetrically} with respect to 
the inter-level distance, $\omega$, i.e.
 $\delta_2=-\delta_1=\delta$.  
In this particular case Eq. \ref{TILDE} takes the form

\begin{equation}
 \label{symmetric}
 -\frac{d^2{{\tilde A}_1}}{dt^2}
-\frac{{\tilde A}_1}{4}\Biggl( \delta^2+16\lambda^2\cos^2\delta t+\frac{\delta^2}{\cos^2\delta t}                \Biggr)=0.
\end{equation}
From the form of Eq. \ref{symmetric} it can be concluded that
the solution is singular in the vicinity of the time moments,
$t_n=\frac{(2n+1)\pi}{2\delta}$, where it can be viewed as a 
Schr{\"o}dinger equation in the attractive potential 
$-\frac{1}{4(t-t_n)^2}$. The form of the singular solution is
 $ {\tilde A}_1\propto (t-t_n)^{1/2}$. Note that, within the RWA,
 Eq. \ref{symmetric} is exact. The only small-time scale, which can
 "cure" the singularity is $\sim \Delta_Z^{-1}\ll \lambda^{-1}, \delta^{-1}$  is related to the violation of the RWA. Certainly,
 any asymmetry, $(\Delta \lambda)$, in the amplitudes of the two drives would cut off the singularity at $(t-t_n)\lesssim \frac{(\Delta \lambda)}{\lambda}$.

\end{document}